\documentstyle[sprocl,epsfig]{article}

\newcommand{\gamp}{\gamma_1}
\newcommand{\gamx}{\gamma_2}

\begin{document}

\title{POWER SPECTRA FOR GALAXY SHAPE CORRELATIONS}
\author{JONATHAN MACKEY}

\address{Harvard-Smithsonian Center for Astrophysics, 60 Garden St., \\
Cambridge, MA 02138, USA\\E-mail: jmackey@cfa.harvard.edu} 

\maketitle\abstracts{ 
It has recently been argued that the observed ellipticities of
galaxies may be determined at least in part by the primordial tidal
gravitational field in which the galaxy formed.  Long-range
correlations in the tidal field could thus lead to an
ellipticity-ellipticity correlation for widely separated galaxies.  
I present results of a calculation of the angular power spectrum of 
intrinsic galaxy shape correlations using a new model relating 
ellipticity to angular momentum.
I show that for low redshift galaxy surveys, the model predicts that
intrinsic correlations will dominate correlations induced by weak
lensing, in good agreement with previous theoretical work and
observations.  The model also produces `$E$-mode' correlations
enhanced by a factor of $3.5$ over `$B$-modes' on small scales, making
it harder to disentangle intrinsic correlations from weak lensing.}

\section{Introduction}
The study of galaxy alignments has a rich history, as documented by
 Djorgovski~\cite{Djo87}, although with mixed results (e.g.~Cabanela \& 
Aldering~\cite{CabAld98}).
Recently, Lee \& Pen~\cite{lp00a} investigated a relationship between
galaxy spins and the underlying gravitational potential field with a
view to reconstructing this potential.  It was quickly realised that
spatial correlations in the gravitational potential could induce
correlations in the spins of nearby galaxies.
This is interesting in its own right, but is also a potential
contaminant of field-surveys for weak gravitational lensing by large
scale structure.

Weak lensing shear, the coherent distortion of galaxy images on the sky 
induced by density perturbations along the line of sight
\cite{mellier99} has now been detected by
several different groups \cite{RhoRefGro01}.
Following standard practice, all of these authors assume that all of
their observed correlation in the ellipticities of galaxies comes
from weak lensing.
Intrinsic shape correlations, if present, should be considered when
interpreting results from these field-lensing surveys.

Several authors have investigated these intrinsic galaxy shape correlations 
recently:
with analytic arguments~\cite{lp00a,ckb00,cnpt00a,cnpt00b,me01}, using
numerical simulations~\cite{HeaRefHey00,CroMet00}, 
and with observations~\cite{pls00,brownetal00}.  
The calculation and results presented below can be found in more detail in
Mackey, White \& Kamionkowski~\cite{me01}.  Our calculation builds on
ideas contained in Catelan, Kamionkowski \& Blandford~\cite{ckb00},
and uses some similar physical assumptions to Crittenden {\em et
al.}~\cite{cnpt00a}.

\section{Ellipticity Model and Power Spectra Calculation}
Here I briefly discuss the model and calculation.  Using the same
formalism as in weak lensing analyses, I describe the intrinsic
ellipticity of galaxies in terms of the spin-2 (complex) ellipticity:
${\bf \epsilon} = |\epsilon|e^{2i\phi} = \gamma_{1}+i\gamma_{2} \;.$
I assume the components $\gamma_i$ of the ellipticity are determined
by a galaxy's angular momentum.  This makes sense in that a rotating
system will become flattened perpendicular to the angular
momentum vector, and will be more flattened for systems with higher
angular momentum.  In this case, as argued by Catelan {\em et
al.}~\cite{ckb00}, the intrinsic ellipticity is given by
\begin{equation}
\gamp = f(L,L_{z}) (L_{x}^{2}-L_{y}^{2}) \; \; {\rm and} \; \;
\gamx = 2 f(L,L_{z}) L_{x}L_{y} \,,
\label{eqn:ellipdef}
\end{equation}
where the sky is the {\em x-y} plane, and where
$f(L,L_{z})$ is an unknown function which determines how ellipticity
scales with $L$.  I take $f(L,L_{z})$ to be a constant, $C$, whose value must
be fitted empirically to the observed rms ellipticity of galaxies.  This 
means that $| \epsilon | \propto L^2$.

Galaxies acquire angular momentum during formation by tidal torques.
It can be shown~\cite{white84}
that the angular momentum acquired to first order in the gravitational
potential $\Phi({\bf x})$, is
$ L_{i} \propto \epsilon_{ijk} I_{kl}\partial_{l} \partial_{j}\Phi({\bf x}) \;,$ 
where $I_{ij}$ is the protogalaxy's inertia tensor.

The ellipticity components $\gamma_i$ can thus be calculated in terms
of the gravitational potential.  I then decompose the spin-2
ellipticity field into scalar (electric-type $E$-mode) and
pseudo-scalar (magnetic-type $B$-mode) fields in Fourier space. 
This enables the
construction of 3D power spectra for the ellipticities, which are
convolutions over the density power spectrum.  Finally,
the Limber approximation is used to obtain the
predicted angular power spectrum for different source galaxy distributions.

\section{Results and Discussion}
\begin{figure}[t]
\epsfig{figure=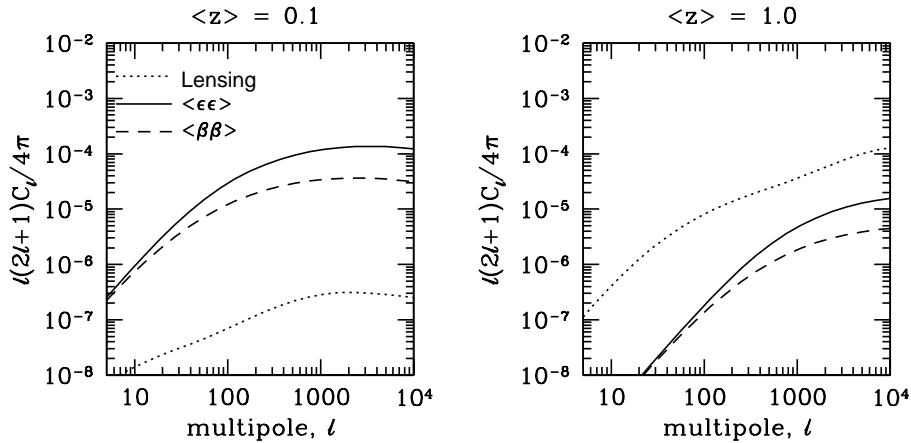,height=2.3in}
\caption{The angular power spectra of intrinsic shape correlations as
predicted by this model.  The solid line is the $EE$, and the dashed
line the $BB$ intrinsic power spectrum.  The $EB$ cross power spectrum
is parity violating, so it is identically zero.  For comparison, the
dotted line is the predicted weak lensing signal.  The left panel is
for a low redshift source distribution with mean $\langle z_{{\rm
src}} \rangle =0.1$, and the right panel a high redshift distribution
with mean $\langle z_{{\rm src}} \rangle = 1.0$.  For reference, $\ell
\sim 200$ corresponds to an angular scale of $\theta \sim 1^{\rm o}$.
\label{fig:l2cl}}
\end{figure}

To demonstrate the results I use a low and a high redshift source 
distribution (with mean source redshifts
$\langle z_{\rm src}\rangle=0.1$ and 1.0).  The ellipticity-ellipticity
angular power spectra obtained are shown in Fig.~\ref{fig:l2cl}, along 
with the corresponding weak lensing prediction.

The most obvious feature in Fig.~\ref{fig:l2cl} is that for a low 
redshift survey, 
intrinsic correlations are expected to dominate over weak lensing signal,
while with 
$\langle z_{\rm src}\rangle=1.0$, they are a very 
small contaminant to weak lensing measurements.  
The reasons for this result are twofold:  the lensing signal is proportional
to the projected density which increases with survey depth, while the 
intrinsic signal is increasingly washed out by projection effects for 
deeper surveys.
The predicted amplitudes are in good agreement with the calculations of Crittenden {\em et al}~\cite{cnpt00a}, and comparable to the observational results of Brown {\em et al}~\cite{brownetal00} over angular scales from $10'-100'$.  They are also comparable to findings from numerical 
simulations~\cite{CroMet00,HeaRefHey00}.

The shape of the angular power spectrum reflects the shape of the 
underlying density power spectrum $\Delta^{2}_{m}(k)$, going roughly flat
on small scales.  On large scales the log-slope is $2$, consistent with shot 
noise.  
The shapes of the intrinsic and lensing power spectra are quite similar, 
making it difficult to use this to separate the two components.

A potentially better discriminant is given by the relative levels of
$E$- and $B$-mode power (this was also investigated by Crittenden {\em
et al.}~\cite{cnpt00b} in their model).  Weak lensing shear has no
handedness and can therefore produce only $E$-modes.
Fig.~\ref{fig:l2cl} shows that the intrinsic power spectra have equal
$E$- and $B$-mode power on large scales, but
$C_{\ell}^{\epsilon\epsilon} / C_{\ell}^{\beta\beta} \simeq 3.5$ on
small scales.  This has two important implications for weak lensing.
First, a detection of $B$-mode power would indicate that intrinsic
correlations are present.  Second, intrinsic correlations can still
hide in a low signal-to-noise measurement of $E$-mode only power (but
only up to a factor of $3.5$), because of this $E$-mode enhancement on
small scales.  Thus, some caution should be used in estimating noise
and contamination of weak lensing measurements solely from the
level of $B$-mode power.

The $E$-mode enhancement can be understood intuitively by noting that
isolated point masses can generate only $E$-modes.  On smaller scales 
the density field can be increasingly described in terms of distinct
objects, giving a mostly $E$-mode signal.  $B$-modes arise because 
the ellipticities $\gamma_{i}$ 
are convolutions over the Fourier 
modes of the potential.  It was found that large scale perturbations 
on the small scale potential field produce the $B$-modes.

The small scale $E$-mode enhancement is a distinctive feature of this
model, and could be tested with current observational data.  It is
qualitatively different from that of Crittenden {\em et
al}.~\cite{cnpt00b}, who found that $E$- and $B$-modes are the same on
small scales and different on large scales.  
The halo shapes ellipticity model of Catelan {\em et al}.~\cite{ckb00} 
is also different in that it produces only $E$-modes.  
Thus, the $E$-$B$ decomposition is potentially a very good
way to observationally distinguish between the different models of
intrinsic correlations.  Determining the correct model may give us new
insight into galaxy formation and evolution processes, so it is
important to do so.

\section*{Acknowledgments}
I thank Martin White and Marc Kamionkowski for collaboration and advice 
on this work.  Thanks especially to MW for many hours of discussion and 
lots of good ideas and guidance.

\end{document}